\documentclass[fleqn,usenatbib]{mnras}

\usepackage{newtxtext,newtxmath}
\usepackage[T1]{fontenc}
\usepackage{ae,aecompl}

\usepackage{graphicx}	% Including figure files
\usepackage{epstopdf}	% Including figure files
\usepackage{color}

\newcommand{\green}[1]{{\color{\green} #1}}

\title[\textit{SRG}/eROSITA view of X-ray reflection in {the} CMZ]{\textit{SRG}/eROSITA view of X-ray reflection in {the} Central Molecular Zone: { a snapshot in September-October 2019}}
\author[Khabibullin, Churazov \& Sunyaev]{
Ildar Khabibullin$^{1,2}$,
Eugene Churazov$^{1,2}$,
and Rashid Sunyaev$^{1,2}$
\\\\
$^{1}$ Space Research Institute, Profsoyuznaya str. 84/32, Moscow, 117997, Russia\\
$^{2}$ MPI f\"ur Astrophysik, Karl-Schwarzschild str. 1, Garching D-85741, Germany\\
}

\date{Accepted XXX. Received YYY; in original form ZZZ}

\pubyear{2021}

\begin{document}
\label{firstpage}
\pagerange{\pageref{firstpage}--\pageref{lastpage}}
\maketitle

\begin{abstract}
X-ray reflection off dense molecular clouds in the Galactic Centre region has established itself as a powerful probe for the past activity record of the supermassive black hole Sgr~A* on a timescale of a few hundred years. Detailed studies of the reflection signal from individual clouds allow us to estimate parameters of the brightest flare(s) and explore properties of the dense gas distribution inside and around them. On the other hand, wide area surveys covering the full Central Molecular Zone (CMZ) are crucial to spot brightening of the new clouds and long-term decay of the flux from the once bright ones. {Here we present and discuss the data obtained by the \textit{SRG}/eROSITA telescope in the course of its commissioning observations in late 2019 in regard to the X-ray reflection off CMZ clouds located to the East of Sgr~A*}. We measure the hard X-ray (4-8 keV) flux from the currently brightest (in reflected emission) molecular complex Sgr A and derive upper limits for other molecular complexes.  We confirm that the Sgr A complex keeps being bright at the level of $4\times10^{-13}\, {\rm erg\,s^{-1}\,cm^{-2}\,arcmin^{-2}}$, making it an excellent candidate for the deep forthcoming high resolution imaging and polarimetric observations. We also discuss distinct features of the reflected emission from the clouds for which the primary illumination front has already passed away and the signal is dominated by multiply scattered radiation. 
\end{abstract}
%%%%%%%%%%%%%%%%%%%%%%%%%%%%%%%%%%%%%%%%%%%%%%%%%%

%%%%%%%%%%%%%%%%% BODY OF PAPER %%%%%%%%%%%%%%%%%%
%------------------------------------------------
\section{Introduction}
\label{s:introduction}
%------------------------------------------------

~~~~~Thanks to the light-travel-time delay, X-ray reflection off dense molecular gas in the Central Molecular Zone (CMZ) of our Galaxy gives us an opportunity to observe 'echoes' of the past X-ray outbursts from the supermassive black hole Sgr~A* in real time \citep[see][for a review]{2013ASSP...34..331P}. Being a couple hundred parsec across, the full CMZ can be 'scanned' by X-ray echo of a single flare over the time-span of {a} few hundred years, allowing one to reconstruct the 3D distribution of the dense molecular complexes with respect to the source of primary emission \citep[e.g.][]{2017MNRAS.468..165C}. Moreover, sensitive observations of the brightest illuminated clouds open a way to explore their interiors down to sub-pc scales {and free from projection effects}, given that duration of the original flare lasted for not more than a few years \citep{2017MNRAS.471.3293C,2020MNRAS.495.1414K}.

So far the X-ray reflection signal has been detected from several molecular complexes, starting from the initial discovery of the hard X-ray emission from the Sgr~B2 molecular complex by \textit{GRANAT}/ART-P \citep[][]{1993ApJ...407..606S}. Such interpretation led to the prediction of the existence of the bright 6.4 keV iron fluorescent line in the scattered radiation spectrum \citep[][]{1993ApJ...407..606S}. This prediction and the ART-P discovery were confirmed by the detection of the fluorescent line emission by \textit{ASCA} \citep[][]{1996PASJ...48..249K} and hard X-ray emission by \textit{INTEGRAL} \citep[][]{2004A&A...425L..49R}.
Later on, variable diffuse X-ray emission featuring hard spectrum and fluorescent line of iron at 6.4 keV has been detected from the Sgr~A complex \citep[][]{2007ApJ...656L..69M,2010ApJ...714..732P,2012A&A...545A..35C,2013A&A...558A..32C,2017MNRAS.465...45C}, parts of Sgr~B \citep[][]{2008PASJ...60S.191N,2009PASJ...61S.241I,2010ApJ...719..143T,2015ApJ...815..132Z} and Sgr~C \citep[][]{2001ApJ...550..297M,2009PASJ...61S.233N,2013PASJ...65...33R,2018A&A...610A..34C} complexes, probably the Sgr~D complex \citep[][]{2018A&A...612A.102T}, the cloud adjacent to the Arches Cluster \citep[][]{2011A&A...530A..38C,2014ApJ...781..107K,2014MNRAS.443L.129C,2017MNRAS.468.2822K,2019MNRAS.484.1627K}, and a few others \citep[see][for a thorough overview of the early results and latest \textit{XMM-Newton} measurements]{2018A&A...612A.102T}.

Thorough characterisation of the refection signal from of the brightest molecular complexes made it possible to measure the key parameters of the flare(s) responsible for the observed echo and construct models of the echo propagation through the CMZ on a time-scale of few hundred years \citep[e.g.][]{2017MNRAS.468..165C}. {Forthcoming deep observations of the Sgr A complex with \textit{Chandra} \citep[][]{2000SPIE.4012....2W} and \textit{IXPE} \citep[][]{2016SPIE.9905E..17W} observatories, aimed {at} the highest resolution imaging \citep[][]{2017MNRAS.471.3293C,2020MNRAS.495.1414K} and X-ray polarimetry \citep[][]{2020MNRAS.498.4379K,2020A&A...643A..52D,2021arXiv210906678F}, respectively, will refine the models even further and bring them to the level where robust characterisation of the molecular clouds themselves will become feasible \citep[see][outlining the long-term road-map for these studies]{2019BAAS...51c.325C}.}

On the other hand, since the line-of-sight location of many other molecular complexes is not known, one cannot firmly predict when they will start being bright in the X-ray reflection. In this regard, it is crucial to map the full CMZ with sufficient sensitivity to spot possible brightening of the diffuse hard X-ray and fluorescent iron emission from their direction.

{Here we present the X-ray data obtained by the eROSITA telescope \citep[][]{2021A&A...647A...1P} onboard \textit{SRG} X-ray observatory \citep[][]{2021arXiv210413267S} during its commissioning phase in late 2019. The presented data cover full extent of the CMZ to the East of Sgr~A*, where the Russian \textit{SRG}/eROSITA consortium is responsible for the scientific data analysis \citep[e.g.][]{2021arXiv210413267S}. Although {only a few of the seven} telescope modules of \textit{SRG}/eROSITA were operating at this moment \citep[][]{2021A&A...647A...1P}, the data are {sufficiently} sensitive and characterised by uniform and stable background.} This allows us to measure robustly the flux from the brightest (in reflected X-ray emission) molecular complex Sgr~A and put constraints on the flux from other complexes, including the once-bright Sgr~B2 complex. Although the X-ray illumination front has already fully scanned the latter cloud several years ago, one might expect the contribution of the multiply-scattered emission to last much longer. We discuss the currently observed signal and future prospects in this regard.   

%newpage
%------------------------------------------------
\section{X-ray Data}
\label{s:data}
%------------------------------------------------
%-----------------------------------------
%%%%
\begin{figure*}
    \centering
    \includegraphics[bb = 50 300 530 550,width=0.85\textwidth]{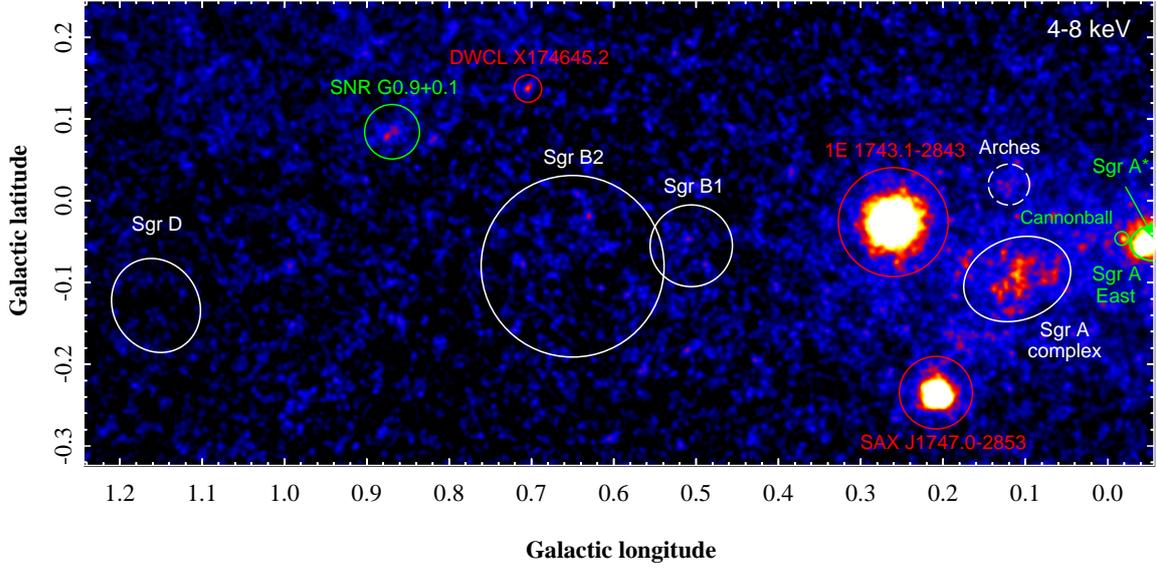}
    \caption{Background-subtracted vignetting-corrected \textit{SRG}/eROSITA image of the "Eastern" Central Molecular Zone in the 4-8 keV band. Positions of massive molecular complexes are marked with white circles, including the cloud adjacent to the Arches cluster. A number of other hard X-ray sources (both diffuse and point-like) are visible and labelled in the image.}
    \label{f:image4080lab}
\end{figure*}
%%%
%-----------------------------------------

The \textit{SRG} observatory \citep[][]{2021arXiv210413267S}, featuring two X-ray telescopes - Pavlinsky ART-XC \citep[][]{2021A&A...650A..42P} and eROSITA \citep[][]{2021A&A...647A...1P},  was launched on July 13, 2019 and started conducting commissioning and performance verification observations a month later, being on the way to its final orbit around the Earth-Sun L2 point. The Galactic Centre region has been selected as one of the first targets for commissioning observations of the ART-XC telescope in August 15-16, 2019, when {the} eROSITA telescope has not yet started its operations. Later on, {a few pointed observations and} a  wide-field survey of the Galactic Centre region have been performed by \textit{SRG} in September-October 2019, now with {one to three} telescope modules {(out of seven)} of eROSITA having been switched on. Here we report the results of these observations at "positive" Galactic longitudes (more precisely, East to the position of Sgr A*), where the data rights were assigned to the Russian \textit{SRG}/eROSITA consortium. Namely, we focus on the data from 4 to 8 keV and investigate X-ray reflection signal from molecular clouds in the Central Molecular Zone of our Galaxy.

Due to the drop of eROSITA's effective area {at the energy of the gold absorption edge above $\sim$2 keV}, the data in the band of consideration is dominated by the background count rate caused by the energetic particles hitting the detector \citep[see][for the details]{2021A&A...647A...1P}. On the other hand, this background signal turned out to be very stable over approximately the first year of eROSITA operation, allowing one to collect enough data for accurate description of its spectral properties, as well as temporal and spatial (over the detector) variations. In what follows, we use the particle background model constructed from the so-called Filter Wheel Closed data, i.e. the data accumulated during the periods when the detectors were shielded from the direct sky exposure by a filter \citep[see][for the details]{2021A&A...647A...1P}. 

After checking for the background flares and subtraction of the model particle background signal, one can correct the data for the energy dependent vignetting of the telescope response. We use a version of the vignetting function calibrated specifically on \textit{SRG}/eROSITA observations of the Coma cluster \citep[][]{2021A&A...651A..41C} and {verified} later on the observation of the UGC~03957 group (Khabibullin et al., in prep). Thanks to the \textit{SRG}'s field-scanning mode  \citep[][]{2021arXiv210413267S} \citep[which has also been employed for a number of Performance Verification observations, e.g.][]{2021A&A...651A..41C,2021arXiv210614517B}, the effective {(i.e. expressed in terms of equivalent on-axis exposure of one eROSITA telescope module after taking into account vignetting and actual number of active telescope modules)} exposure map of the field is rather uniform and exceeds {4 ks} at every point, reaching approximately twice larger values within $\sim15$ arcmin of Sgr~A*, where one of the pointing {\citep[including 'parking' pointings needed in the scanning mode of the spacecraft operation for the data downlink, e.g.][]{2021arXiv210413267S}} observations contributes a sizeable fraction of the total exposure. Roughly similar (vignetting-corrected) exposure times are expected to be reached for this region after completion of at least 4 all-sky surveys, assuming nominal operation of all 7 telescope modules. Thus, the data we present here offer a suitable baseline for the future comparison with the all-sky data as well.

The resulting background-subtracted exposure-corrected map of the surface brightness in {the} 4-8 keV band is shown in Fig.~\ref{f:image4080lab}, with the locations of brightest point sources, supernova remnants, and massive molecular complexes being marked. The image is 1.3~deg$\times$0.5 deg on a side, and covers the full CMZ to the East of Sgr~A*, containing such prominent molecular complexes as Sgr~A, Sgr~B1, Sgr~B2, Sgr~D, Arches cloud, and the Brick Cloud {(G0.253+0.016)} \citep[e.g.][]{2012MNRAS.419.2961J,2020ApJS..249...35B}. The latter one, however, {is located at $(l,b)\approx(0.25\deg,0.0\deg)$ and} projects in the vicinity of very bright point source, 1E~1743.1-2843 \citep[e.g.][]{2016ApJ...822...57L,2019MNRAS.485.2457H}, which makes it very challenging to study potential faint hard diffuse emission coming from it. Even though 4-8 keV band is not the band of primary sensitivity for eROSITA, one can easily notice diffuse emission coming from the Sgr~A molecular complex and Arches cluster, as well as supernova remnants SNR G0.9+0.1 \citep[][]{2003A&A...401..197P} and Sgr~A East with the Cannonball \citep[][]{2013ApJ...778L..31N}. The other prominent sources visible in this band are SAX J1747.0-2853  and a dust-enshrouded, late carbon-type Wolf-Rayet binary DWCL~X174645.2 \citep[][]{2010ApJ...710..706M}. Discussion of their properties lies beyond the scope of the current paper, and we simply mask the regions of their noticeable impact in further analysis.   
%\citep[e.g.][]{2016ApJ...825..132H}
%------------------------------------------------
\section{X-ray reflection signal}
\label{s:reflection}
%------------------------------------------------

%-----------------------------------------
%%%%
\begin{figure*}
    \centering
    \includegraphics[bb = 50 300 530 550,width=0.85\textwidth]{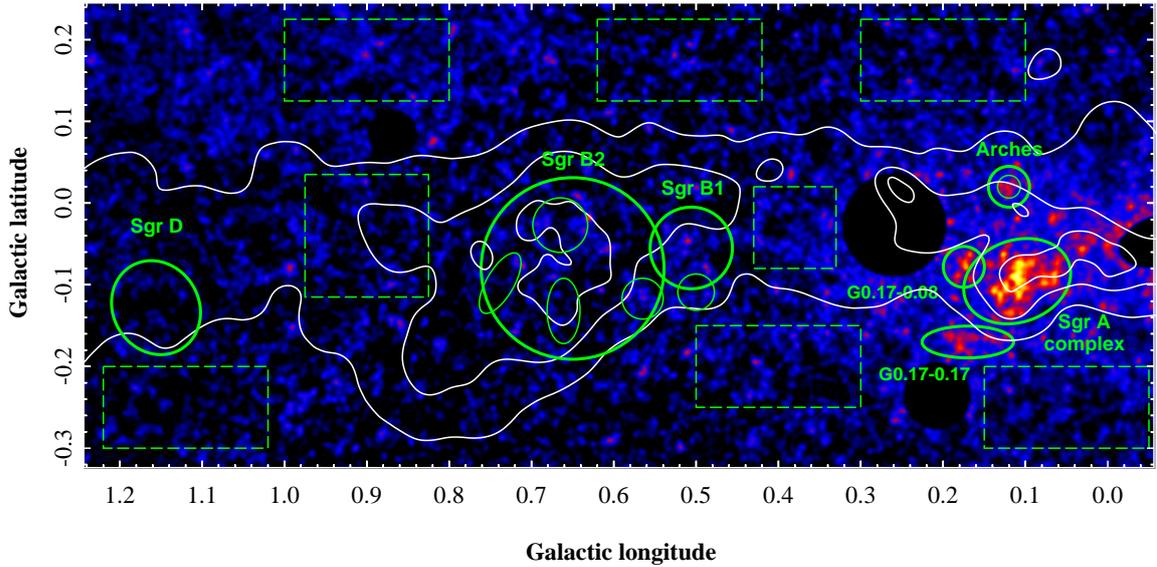}
    \caption{{SRG/eROSITA image in the 4-8 keV band after masking of bright point sources and supernova remnants. White contours show the distribution of molecular gas emission in N$_2$H+ line from the Mopra survey of the CMZ \citep[][]{2012MNRAS.419.2961J}. The solid circles and ellipses show regions used for estimation of the flux from the indicated molecular complexes ({listed} in Table~\ref{t:table_v0}), while the thin dashed boxes show regions used for estimation of the background emission.}}
    \label{f:image4080reg}
\end{figure*}
%%%
%-----------------------------------------

%%%%
\begin{table*}
    \centering
    \begin{tabular}{cccccccccc}
\hline
  \multicolumn{1}{c}{Name} &
  \multicolumn{1}{c}{$l$} &
  \multicolumn{1}{c}{$b$} &
  \multicolumn{1}{c}{R$_1$} &
  \multicolumn{1}{c}{R$_2$} &
  \multicolumn{1}{c}{PA} &
  \multicolumn{1}{c}{SB$_{4-8}$}&
  \multicolumn{1}{c}{SB$_{\rm bs,4-8}$} &
  \multicolumn{1}{c}{SB$_{\rm bs,6.4}$} &
  \multicolumn{1}{c}{Flux$_{4-8}$}
   \\
   \multicolumn{1}{c}{} &
  \multicolumn{1}{c}{deg} &
  \multicolumn{1}{c}{deg} &
  \multicolumn{1}{c}{arcsec} &
  \multicolumn{1}{c}{arcsec} &
  \multicolumn{1}{c}{deg} &
  \multicolumn{1}{c}{erg/s/cm$^2$/arcmin$^2$}&
  \multicolumn{1}{c}{erg/s/cm$^2$/arcmin$^2$} &
  \multicolumn{1}{c}{ph/s/cm$^2$/arcmin$^2$} &
  \multicolumn{1}{c}{erg/s/cm$^2$}
  \\
\hline
 
  Sgr A & 0.110 & -0.096 & 240 & 180 & 20 & $(46\pm2)\times 10^{-14}$ & $(37\pm3)\times 10^{-14}$ &$(62\pm13)\times 10^{-7}$ &$(138\pm9)\times 10^{-13}$\\
 G0.17-0.08 & 0.17 & -0.08 & 90 & 90 & 0 & $(30\pm4)\times 10^{-14}$ & $(22\pm4)\times 10^{-14}$&$(50\pm18)\times 10^{-7}$ & $(22\pm4)\times10^{-13}$\\
 \smallskip
  G0.17-0.17 & 0.17 & -0.17 & 200 & 70 & 0 & $(24\pm3)\times 10^{-14}$ & $(20\pm4)\times 10^{-14}$&$(44\pm13)\times 10^{-7}$ & $(24\pm4)\times10^{-13}$\\
   
   Arches 50''& 0.120 & 0.020 & 50 & 50 & 0 & $(45\pm10)\times 10^{-14}$ & $(35\pm10)\times 10^{-14}$ & $(10\pm20)\times 10^{-7}$& $(7\pm3)\times10^{-13}$\\
  Arches 100''& 0.120 & 0.020 & 90 & 90 & 0 & $(31\pm4)\times 10^{-14}$ & $(22\pm4)\times 10^{-14}$ & $(17\pm16)\times 10^{-7}$& $(16\pm3)\times10^{-13}$\\
  Sgr B1 & 0.509 & -0.055 & 180 & 180 & 0 & $(12\pm2)\times 10^{-14}$ & $(4\pm2)\times 10^{-14}$&$(10\pm9)\times 10^{-7}$ & $(11\pm6)\times10^{-13}$\\
  G0.50-0.11 & 0.500 & -0.109 & 80 & 80 & 0 & $(10\pm4)\times 10^{-14}$ & $(3\pm4)\times 10^{-14}$&$(0\pm17)\times 10^{-7}$ & $(2\pm2)\times10^{-13}$\\
    G0.56-0.11 & 0.565 & -0.117 & 90 & 90 & 0 & $(12\pm4)\times 10^{-14}$ & $(7\pm4)\times 10^{-14}$&$(50\pm15)\times 10^{-7}$ & $(5\pm3)\times10^{-13}$\\
  Sgr B2 & 0.650 & -0.08 & 410 & 410 & 0 & $(7\pm1)\times 10^{-14}$ & $(1\pm1)\times 10^{-14}$& $(11\pm6)\times 10^{-7}$ &$(5\pm20)\times10^{-13}$ \\
  Sgr B2 core & 0.665 & -0.027 & 120 & 120 & 0 & $(7\pm2)\times 10^{-14}$ & $(2\pm3)\times 10^{-14}$&$(7\pm11)\times 10^{-7}$ & $(2\pm3)\times10^{-13}$\\
    G0.66-0.13 & 0.661 & -0.132 & 144 & 72 & 90 & $(3\pm3)\times 10^{-14}$ & $(0\pm3)\times 10^{-14}$&$(9\pm13)\times 10^{-7}$ & $(0\pm3)\times10^{-13}$\\
     G0.74-0.10 & 0.500 & -0.109 & 150 & 60 & 60 & $(5\pm3)\times 10^{-14}$ & $(0\pm3)\times 10^{-14}$&$(0\pm12)\times 10^{-7}$ & $(0\pm2)\times10^{-13}$\\
     
  Sgr D & 1.156 & -0.128 & 213 & 188 & 120&$(3\pm1)\times 10^{-14}$ & $(0\pm1)\times 10^{-14}$& $(5\pm6)\times 10^{-7}$ &$(0\pm4)\times10^{-13}$\\
\hline\end{tabular}

    \caption{Parameters of the extraction regions, measured 4-8 keV flux and surface brightness for the most prominent molecular clouds in the covered region compared to the estimated local level of the background emission. The regions are ellipses with the centre at Galactic coordinates ($l,b$), major and minor semi-axes equal to R$_1$ and R$_2$, and the orientation given by the position angle PA (relative to orientation when the major axis is directed to the Galactic West).The SB$_{4-8}$,SB$_{\rm bs,4-8}$ columns give the total and background-subtracted surface brightness of the emission in the 4-8 keV, while Flux$_{4-8}$ gives background-subtracted flux from the whole region. SB$_{\rm bs,6.4}$ is the background-subtracted photon surface brightness in the 6.4 keV line. }
    \label{t:table_v0}
\end{table*}
%%%%

Since we are interested in the signatures of X-ray reflection off molecular clouds within CMZ, we first mask point-like sources (including their possible dust-scattering halos) and unrelated mildly extended sources, and then smooth the image to increase visibility of  largely extended structures. The resulting image is shown in Fig.~\ref{f:image4080reg}, where the white contours outline  distribution of molecular gas emission in N$_2$H+ line from the Mopra survey \citep[][]{2012MNRAS.419.2961J}. As we noticed earlier, the region of the Brick Cloud \citep[G0.253+0.016][]{2019MNRAS.485.2457H}, located at $(l,b)\approx(0.25\deg,0.\deg)$ gets masked and will be excluded from the further consideration. One can easily notice diffuse emission coming from the Sgr~A complex and the Arches cluster, while for the other clouds, including once-bright Sgr~B2 complex, the signal appears to be hardly distinguishable from the background. We quantify this next via spectral analysis of the data extracted from individual regions similar to those analysed in \cite{2018A&A...612A.102T} {(see e.g. their Table~A.2)} and listed in Table~\ref{t:table_v0}.
 
Given relatively low count statistics of the available data, we fit the extracted spectra for the source and background regions with a phenomenological model consisting of three components absorbed by the same column density, namely written as \texttt{wabs*(\texttt{APEC}+pow+gauss)} in the spectral fitting package \texttt{XSPEC} \citep[][]{1996ASPC..101...17A}. We fix the absorbing column density at $N_H=7\times10^{22}$ cm$^{-2}$, the temperature and metallicity of the thermal \texttt{APEC} component at $kT=$6.5 keV and $Z=1$ (in Solar abundance units), the slope of the powerlaw component at $\Gamma=2$, and the central energy of the gaussian line $E_0=6.4$ keV (with no additional broadening applied).  The thermal component offers a good description for the hard seemingly-diffuse emission which is ubiquitously observed across the Galactic Centre region. The power law component supplemented by the gaussian line at 6.4 keV represents possible contribution of the hard diffuse emission arising from X-ray reflection by the molecular gas \citep[e.g.][]{2010ApJ...714..732P}.

Although a number of physically-motivated models describing X-ray reflection exist \citep[e.g.][]{2017MNRAS.468..165C}, the quality of the eROSITA data in the 4-8 keV band would not allow us to exploit their full diagnostic power. Instead, the phenomenological model used here offers a simple and sufficiently flexible unified description for both the source and background regions \citep[][]{2018A&A...612A.102T}.  

For every source region, we define three nearby background regions shown as dashed boxes in Fig. \ref{f:image4080reg}. From them we estimate the averaged background level both in terms of the total surface brightness in {the} 4-8 keV and photon surface brightness in the 6.4 keV line. The resulting total and background-subtracted values, as well as integrated 4-8 keV flux  from each region are listed in Table~\ref{t:table_v0}. {As one can see from Fig.~\ref{f:image4080reg} and Table~\ref{t:table_v0}, significant non-zero residual signal is detected from the direction of the Arches cluster, Sgr~A molecular complex, and two other regions: G0.17-0.17 and G0.17-0.08 (see Table \ref{t:table_v0} for the exact region parameters). }

\begin{figure}
    \centering
    \includegraphics[bb = 60 50 700 540,width=0.8\columnwidth]{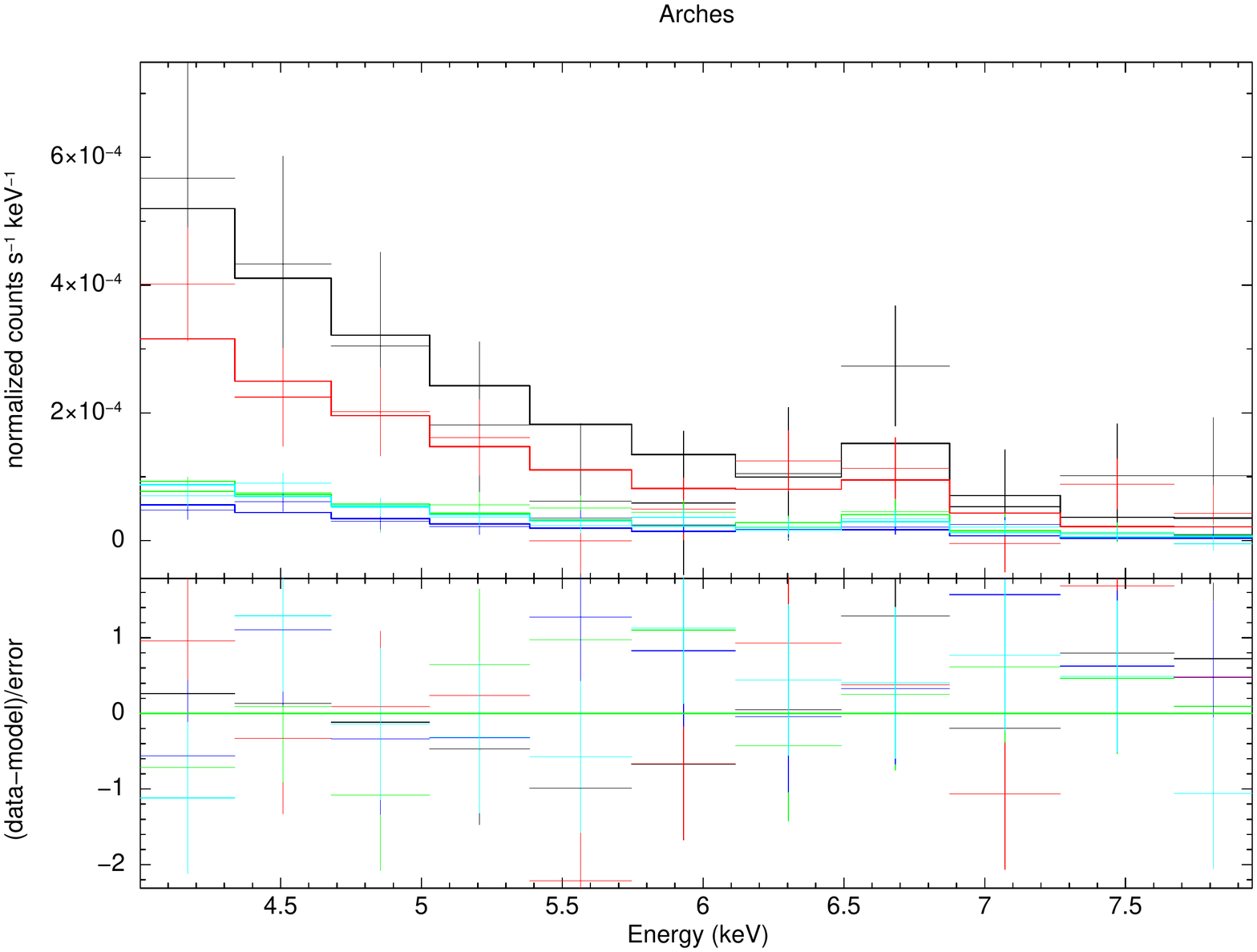}
    \includegraphics[bb = 60 50 700 580,width=0.8\columnwidth]{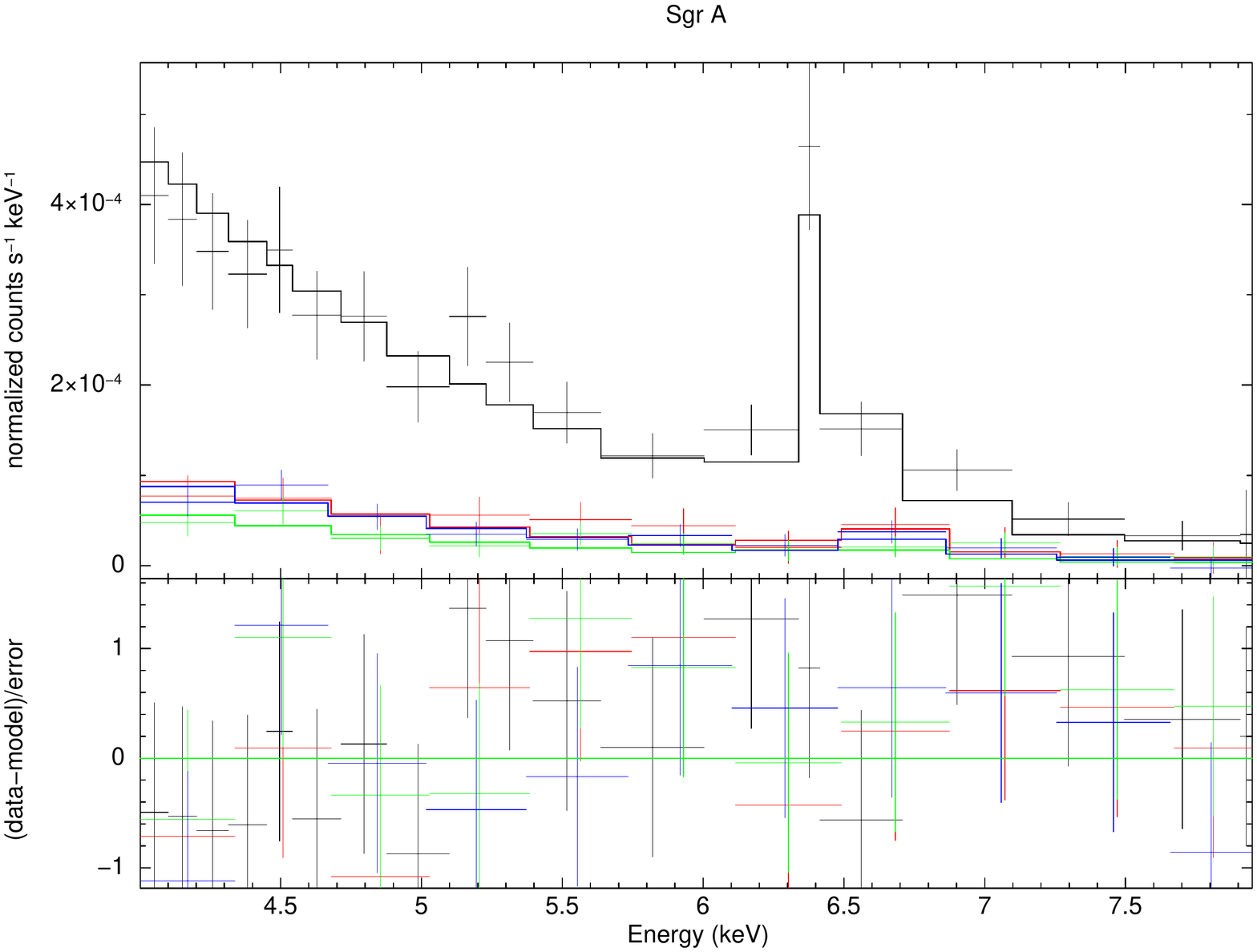}
    \includegraphics[bb = 60 50 700 580,width=0.8\columnwidth]{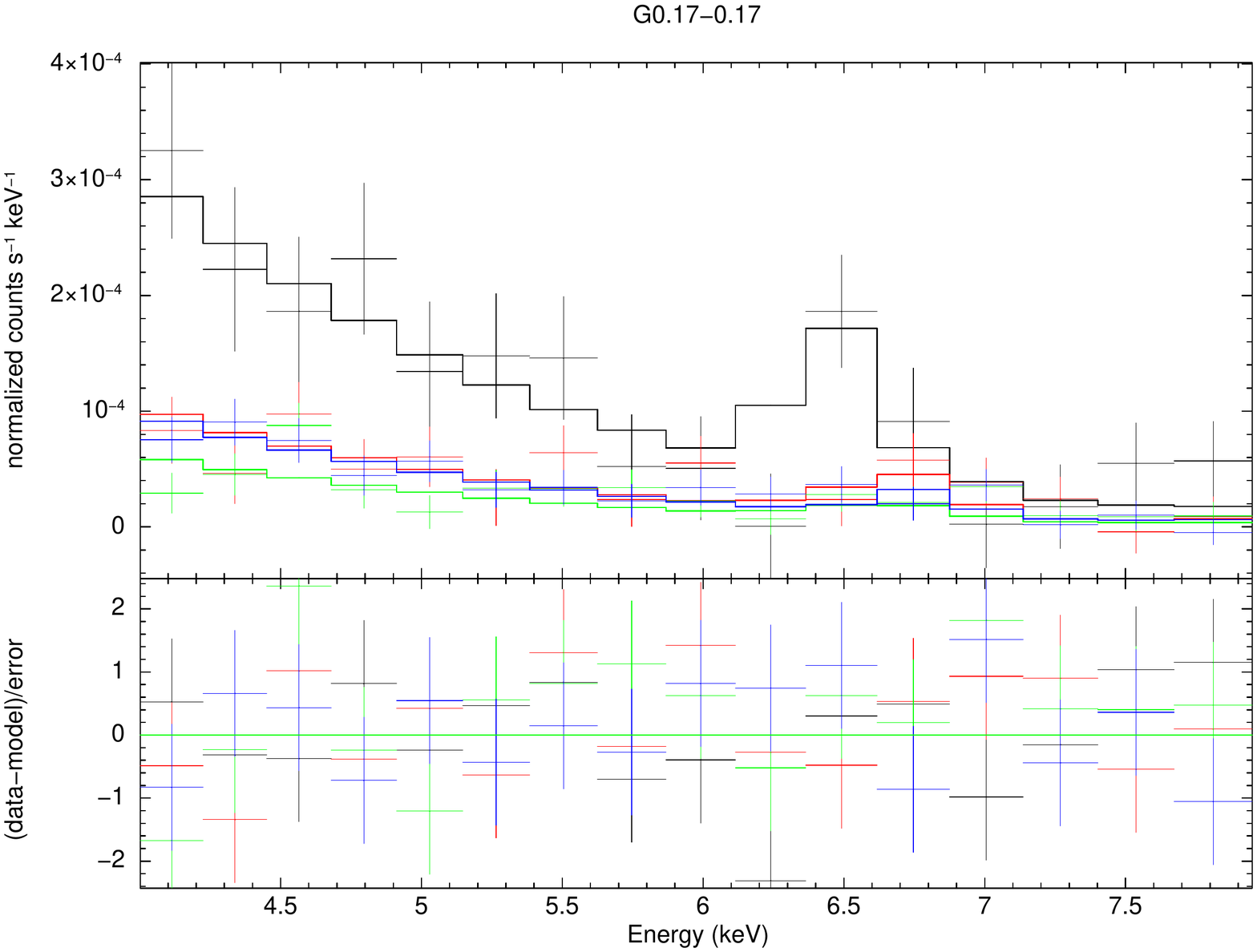}
        \includegraphics[bb = 60 50 700 580,width=0.8\columnwidth]{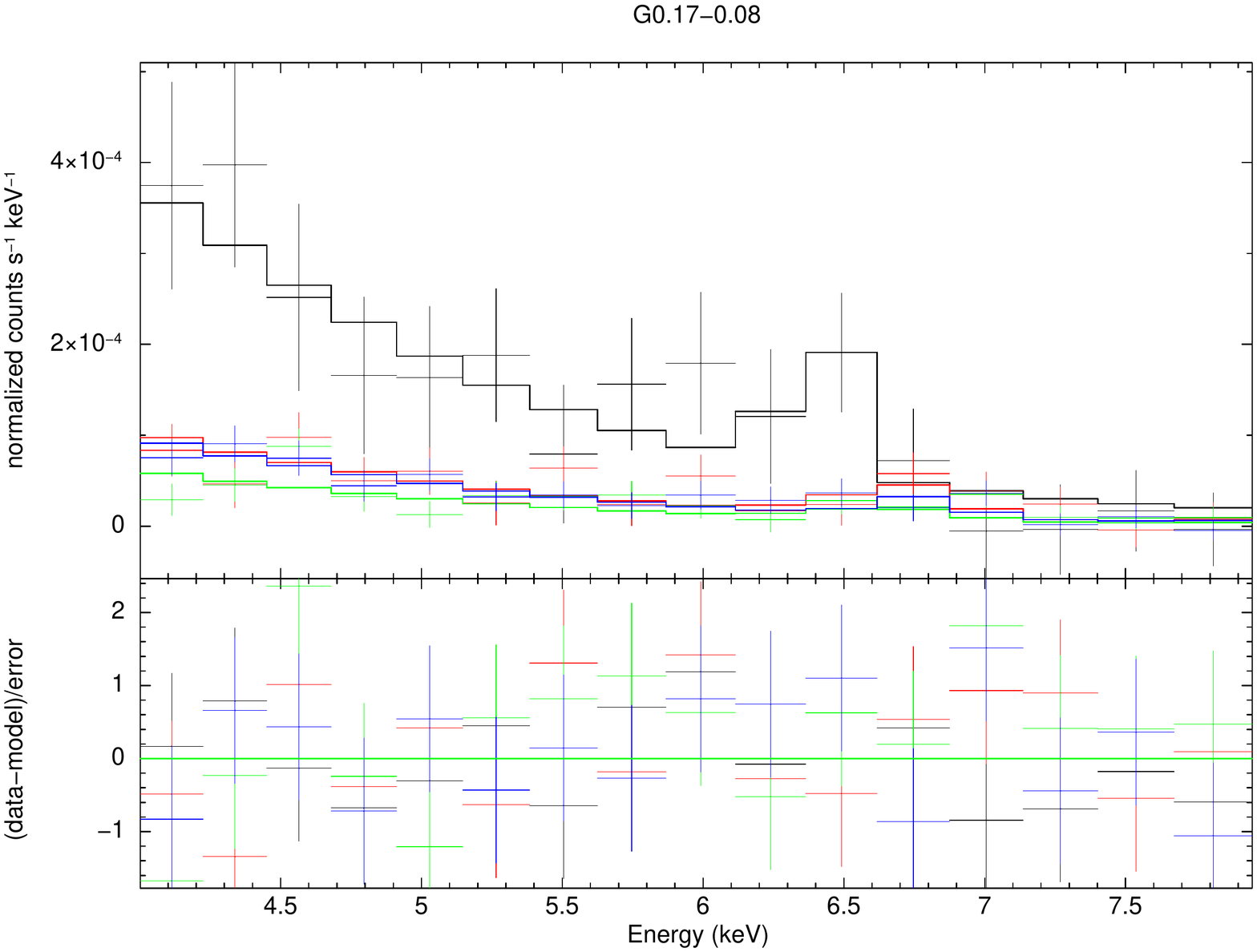}
    \caption{4-8 keV band particle-background-subtracted spectrum of the emission coming from the Arches cloud (top panel, black for R=50'' and red for R=100''), the Sgr~A complex (second panel, black), G0.17-0.17 (third panel, black) and G0.17-0.0.08  (bottom panel, black), in comparison to the spectra from three background regions nearest them (shown in other colours). All spectra are fit with the three-component model described in text (solid lines).}
    \label{f:archessgra}
\end{figure}

{
For the region of {the} Arches Cluster, however, contribution of the hard X-ray emission from the cluster itself plays a dominant role and cannot be reliably filtered out based on the relatively shallow eROSITA data. The total emission is concentrated mostly within a region of 50'' in radius \citep[e.g.][]{2006MNRAS.371...38W,2011A&A...530A..38C,2014MNRAS.443L.129C,2019MNRAS.484.1627K}.  We measure surface {brightness} of the {background-subtracted} 4-8 keV signal at the level of $\approx(3.5\pm1)\times10^{-13}$ erg s$^{-1}$ cm$^{-2}$ arcmin$^{-2}$, which is close to but slightly higher than the latest reported measurement  by \cite{2019MNRAS.484.1627K}, who also used R=50'' region centred on the Arches cluster and analysed data obtained in 2015 and 2016. Conversion of their model (see Table 2 there) into the 4-8 keV surface brightness gives  $\approx(2.0\pm0.2)\times10^{-13}$ erg s$^{-1}$ cm$^{-2}$ arcmin$^{-2}$. The spectrum of the emission is dominated by the thermal component with the brightest iron line at 6.7 keV (see red points in the top panel of Fig.~\ref{f:archessgra}). We detect no significant 6.4 keV line emission from this region, with the 1$\sigma$ upper limit on its photon surface brightness  being {$\lessapprox3\times10^{-6}$ {photons} s$^{-1}$ cm$^{-2}$ arcmin$^{-2}$}, which is fully consistent with their measurement at the level of $\approx(1.1\pm0.5)\times10^{-6}$ erg s$^{-1}$ cm$^{-2}$ arcmin$^{-2}$. If the 100''-radius extraction region is used instead \citep[e.g.][]{2018A&A...612A.102T}, {the} 6.4 keV line emission is only marginally detected with the similar upper limit {$\lessapprox3\times10^{-6}$ {photons} s$^{-1}$ cm$^{-2}$ arcmin$^{-2}$} obtained.

Thus, we conclude that no strong brightening of the reflected X-ray emission from the Arches complex is observed, in agreement with the suggestion that the fading phase of the reflected emission was witnessed over the past decade \citep[][]{2014MNRAS.443L.129C,2017MNRAS.468.2822K,2019MNRAS.484.1627K}.
The slight increase in the emission of the cluster itself might be related to {a flux increase} (by $\approx4\times 10^{-13}$ erg s$^{-1}$ cm$^{-2}$ in 4-8 keV, corresponding to $\approx3\times10^{33}$ erg s$^{-1}$ at the distance of 8 kpc) from any of the numerous points sources contained in it \citep[e.g.][]{2006MNRAS.371...38W}. Future sensitive observation with the arcsec scale angular resolution accessible with the \textit{Chandra} observatory will allow to check this.}

In contrast to this, the reflection signal from the Sgr~A complex is clearly detected and this complex clearly remains the brightest molecular complex in X-ray reflection over the full portion of the CMZ considered here. We measure the background-subtracted 4-8 keV surface brightness at the level of $\approx4\times10^{-13}$ erg s$^{-1}$ cm$^{-2}$ arcmin$^{-2}$ and total flux {(within 4 arcmin by 3 arcmin, i.e. $\approx38$ arcmin$^2$, elliptical region, cf. Figure~\ref{f:image4080reg} and Table~\ref{t:table_v0})} $\approx1.4\times10^{-11}$ erg s$^{-1}$ cm$^{-2}$ (see Table \ref{t:table_v0} for the exact values with corresponding 1$\sigma$ uncertainties). The photon surface brightness in the 6.4 keV line is estimated at the level of $\approx6\times10^{-6}$ photons s$^{-1}$ cm$^{-2}$ arcmin$^{-2}$, fully in agreement with a number of recent measurements for this region \citep[][]{2017MNRAS.465...45C,2017MNRAS.471.3293C,2018A&A...612A.102T}. With the currently presented measurement, being the most up-to-date currently available, we can confirm that the Sgr~A complex indeed stays bright enough to serve as a primary target for the future cutting-edge studies involving high resolution imaging with \textit{Chandra} \citep[][]{2019BAAS...51c.325C,2020MNRAS.495.1414K} and first X-ray imaging polarimetric observation by {\textit{IXPE} \citep[e.g.][]{2019BAAS...51c.325C,2020A&A...643A..52D,2020MNRAS.498.4379K,2021arXiv210906678F}.}

\begin{figure*}
    \centering
    \includegraphics[bb = 75 320 520 520,width=0.9\textwidth]{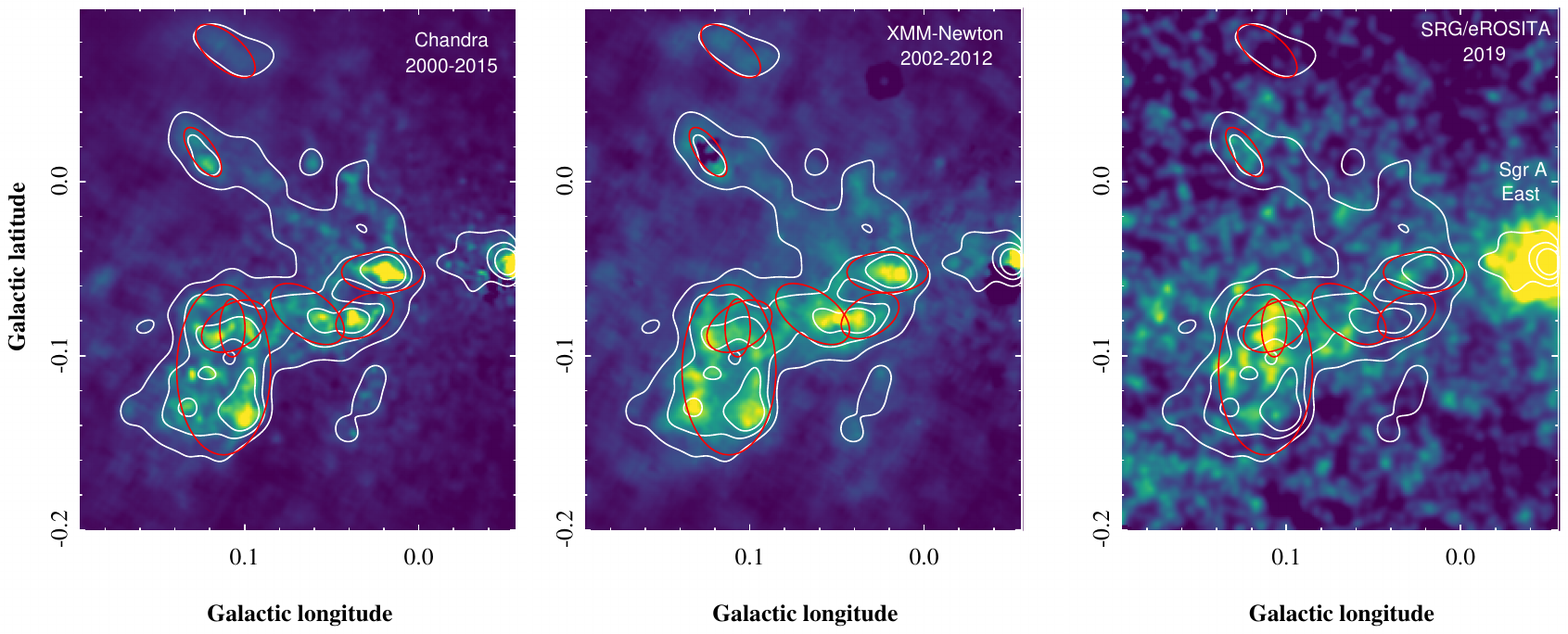}
    \caption{Comparison of the reflected emission from Sgr A molecular complex as observed by \textit{Chandra} (left panel), \textit{XMM-Newton} (middle panel) and {total 4-8 keV emission by \textit{SRG}/eROSITA (right panel).} The white contours are the same for all panels and show distribution of the surface brightness on the \textit{Chandra} image. Regions of the Sgr A complex that have been bright at different epochs in {the} past are shown by red ellipses.}
    \label{f:chandraxmmsrge}
\end{figure*}

The reflected emission from the Sgr~A complex is also characterised by the sophisticated and time-variable morphology \citep[][]{2010ApJ...714..732P,2013A&A...558A..32C,2017MNRAS.465...45C,2018A&A...612A.102T}. We show the morphology of {its} 4-8 keV emission 
%from it 
observed by \textit{SRG}/eROSITA in Fig. \ref{f:chandraxmmsrge} (right panel), in comparison with the reflected emission maps based on archival data of \textit{Chandra} and \textit{XMM-Newton} \citep[left and middle panels, respectively,][]{2017MNRAS.465...45C,2017MNRAS.468..165C}. We overlay regions considered individually by \cite{2018A&A...612A.102T} as red ellipses in Fig. \ref{f:chandraxmmsrge}. One can spot, that the two regions nearest to Sgr~A* in projection, MC1 and MC2, are much fainter in the eROSITA image compared to the \textit{Chandra} and \textit{XMM-Newton} images. On the other hand, the region of so-called "Filament 2011" \citep[][]{2013A&A...558A..32C}, {which was found to be bright in observations taken in 2011 and then was dimming out \citep[][]{2018A&A...612A.102T},} appears to be significantly brighter {(its 4-8 keV surface brightness it is a factor of 1.5 higher than average for the whole cloud)}. The observed rapid {variability} of such extended structures confirms {the} shortness of primary illuminating flare \citep[][]{2013A&A...558A..32C,2017MNRAS.465...45C}, while re-brightening of the "Filament 2011", might be considered as a clear indication in favour of the scenario involving illumination by at least two consecutive flares \citep[][]{2013A&A...558A..32C,2018A&A...612A.102T}. {Alternatively, this filament could be an extended cloud elongated over our line-of-sight.}

This emphasises importance of regular monitoring of the reflected emission, and more sensitive dedicated observation of the Sgr~A complex by \textit{XMM-Newton} and \textit{Chandra} will allow us to verify either of the illumination scenarios.

{There are two regions, adjacent to the Sgr~A complex from which significant 4-8 keV and {the} 6.4 keV line emission is also detected - G0.17-0.08 and G0.17-0.17 (their spectra are shown in the bottom two panels in Fig.~\ref{f:archessgra}). The latter one is essentially the Western half of {the} 400''x70'' filament-like emission region G0.24-0.17 reported by \cite{2018A&A...612A.102T}. We measure background-subtracted surface brightness in {the} 6.4 keV line at the level of $\approx4\times10^{-6}$ photons s$^{-1}$ cm$^{-2}$ arcmin$^{-2}$, which is a factor of 3 higher {than} the value reported by \cite{2018A&A...612A.102T} for the whole G0.24-0.17 region. In the same time, the Eastern part of the G0.24-0.17 filament does not appear as bright on the eROSITA images as it was in the 2012 data presented by \cite{2018A&A...612A.102T}, in line with the rapid echo propagation from the East to the West of the structure suggested by \cite{2018A&A...612A.102T}.

Very similar parameters of the emission are obtained for the region that we call G0.17-0.08 here and which has been briefly discussed in end of the Section 4.6 of  \cite{2018A&A...612A.102T}. According to their Figure~14, the surface brightness in {the} 6.4 keV line was measured at the level of $(2-3)\times10^{-6}$ photons s$^{-1}$ cm$^{-2}$ arcmin$^{-2}$  for this region in 2012, which is a factor of 2 smaller than the measurement by eROSITA in 2019, $(5\pm1.5)\times10^{-6}$ photons s$^{-1}$ cm$^{-2}$ arcmin$^{-2}$. That means we marginally detect a slight increase in the flux from it, although, due to surface brightness variations across the image, an accurate comparison can be done only with the properly matched regions. Thanks to proximity to {the} Sgr~A complex, this region will also benefit from deeper \textit{Chandra} observations to be performed in 2021-2022. }

{For the extended regions centred on Sgr~B1, Sgr~B2 and Sgr~D, we obtain formal {(statistical)} 1$\sigma$ upper limits on the {background-subtracted} 4-8 keV surface brightness  at the levels of { $\lessapprox6\times10^{-14}$ erg s$^{-1}$ cm$^{-2}$ arcmin$^{-2}$, $\lessapprox2\times10^{-14}$ erg s$^{-1}$ cm$^{-2}$ arcmin$^{-2}$ and $\lessapprox1\times10^{-14}$ erg s$^{-1}$ cm$^{-2}$ arcmin$^{-2}$}, respectively. The corresponding limits on the 6.4 keV photon surface brightness are {$\lessapprox19\times10^{-7}$ photons s$^{-1}$ cm$^{-2}$ arcmin$^{-2}$, $\lessapprox17\times10^{-7}$ photons s$^{-1}$ cm$^{-2}$ arcmin$^{-2}$ and $\lessapprox11\times10^{-7}$ photons s$^{-1}$ cm$^{-2}$ arcmin$^{-2}$.}

We have also considered smaller sub-regions of the Sgr~B complex which were analysed by \cite{2018A&A...612A.102T} in order to facilitate the direct comparison. Those include G0.50-0.11, G0.56-0.11, the Sgr~B2 core (G0.66-0.03), G0.66-0.13, and G0.74+0.10 (see Table~\ref{t:table_v0} for the exact parameters). None of them exhibit significant excess over the background level, with the only exception of G0.56-0.11 for which marginally significant excess emission in a narrow band near {the} 6.4 keV line is found. Given that in the broader 4-8 keV band, the statistical significance of the excess turns out to be lower, no firm conclusion can be made on the presence and/or variability of the reflection signal from this region.

For all these complexes we conclude that no strong brightening compared to the values reported by \cite{2018A&A...612A.102T} was observed, although the light curves obtained for some of them based on the \textit{XMM-Newton} monitoring implied possible onset and rise of the fluorescent emission.} The upper limit obtained for Sgr~B2 is also broadly consistent with the latest measurement, implying its transition from being very bright in the reflected emission before 2002 \citep[e.g.][]{2004A&A...425L..49R,2010ApJ...719..143T} to the close-to-background level after 2012 \citep[e.g.][]{2018A&A...612A.102T}. In the next section we briefly discuss implications of all these findings for the X-ray reflection paradigm in general.

%%%%
%%%

%%%%
%%%

%\clearpage
%------------------------------------------------
\section{Discussion}
\label{s:discussion}
%------------------------------------------------

In the short flare paradigm, the loci of clouds being illuminated at any given moment {are} well defined and correspond to a thin paraboloid with the primary source located in focus (more accurately, it is an ellipsoid with both primary source and the observer located in its foci) and thickness along the line-of-sight of $\lesssim 1$ pc for the flare duration less than a few years. In this regard one can distinguish several main phases of illumination for any individual cloud. First, the reflected surface brightness rises due to increase of the mass of the illuminated molecular gas as the front moves from outskirts to the centre of the cloud. Then, the emission stabilises at the level corresponding to the averaged density of the central dense part of the cloud, but with possible significant variations in morphology of emission due to clumpy and filamentary inner structure of the cloud. As the illumination front leaves the dense central part of the cloud, the emission starts to decline and is characterised by more absorbed spectral shape. Finally, at some point, {the} contribution of the multiple scatterings might take over and dominate the observed reflection signal from a cloud. The actual evolution for any individual cloud is primarily determined by its size, mean density and disposition relative to the primary source \citep[see the detailed modelling and discussion in][]{1998MNRAS.297.1279S,2011ApJ...740..103O}. {We note in passing, that even in the absence of any flares, there will be some fluorescent emission due to exposure of molecular gas to {bright nearby X-ray sources \citep[e.g.][]{1993ApJ...407..752C}}, as well as the X-ray \citep[e.g.][]{1980SvAL....6..353V} and cosmic ray backgrounds \citep[e.g.][]{2018ApJ...863...85C}.}

The  Sgr~A complex appears to be in the second stage of this evolution, offering {the} best opportunities to study inner structure and statistical properties of the molecular gas inside it in great detail \citep{2017MNRAS.471.3293C,2020MNRAS.495.1414K}. The presented observations fully confirm this and emphasise benefits of regular observations maximising sensitivity to flux variations on relatively small scales (exemplified by "Filament 2011" structure). 

For a few clouds it was suggested that they might be entering the first stage of illumination. We cannot confirm further significant brightening for Sgr~B1 and Sgr~D to the levels comparable to the Sgr~A or Sgr~B2 complexes in the past. According to the scheme outlined above, observations of the reflected emission on the rising phase are particularly important for exploration of less dense envelopes of the molecular clouds, since no strong contribution from the multiple scattering might be expected at this stage. Further sensitive and uniform observations of the CMZ with \textit{XMM-Newton}, \textit{SRG}/eROSITA and  \textit{SRG}/ART-XC will play a crucial role in finding possible indications of such a brightening. 

Finally, the last two stages are probably taking place now for the Arches and Sgr~B2 cloud. There is a significant difference between these two clouds, however, in terms of the exact behaviour of the declining emission. Indeed, the Arches cloud is likely relatively compact and not very massive, compared to the Sgr~B2 cloud which extends for more than 10 pc and has a mass $\gtrsim$few$\times 10^6$ Solar masses \citep[e.g.][]{2016A&A...588A.143S}. While in the former case one might expect relatively small impact of the multiple scatterings on the late-time light curve, for the latter one their contribution should be present for decades after the illumination front has left the central part of the cloud. 

A number of distinct predictions arise in this case regarding the spectral shape of the observed emission, as described in Appendix. Namely, in addition to the increase in equivalent width of the fluorescent line, one might expect a shift in its mean energy due to enhanced contribution of its Compton shoulder. Although future observations with microcalorimetric resolution will be able to fully resolve the fluorescent line complex, the line energy shift in the multiply scattered emission might already be checked with the spectral resolution provided by \textit{XMM-Newton} (see an illustration in Appendix).

Another interesting implication of the multiple scatterings scenario is {the} virtual independence of the spectral shape above 20 keV upon the incidence angle of the primary emission (unless the cloud is characterised by strong asymmetry in the dense gas distribution) {due to smearing out} of the recoil signatures after a few scatterings. This opens another possibility to check whether different clouds were illuminated by the same flare or whether the spectrum of primary emission was direction-dependent. Since the multiply scattered emission is expected to {stay present} for relatively longer periods of time, accumulation of the signal over years by such missions as \textit{INTEGRAL} and \textit{NuSTAR} might offer an opportunity to check this prediction very soon {\citep[e.g.][]{2021arXiv211007401K}}. 

%------------------------------------------------
\section{Conclusion}
\label{s:conclusion}
%------------------------------------------------

We presented the data of {a} uniform survey of the Galactic Centre region performed by the \textit{SRG} observatory during calibration and performance verification phase in late 2019, focusing on the 4-8 keV data obtained by \textit{SRG}/eROSITA telescope and possible signatures of X-ray reflection off molecular clouds.

We find that Sgr~A complex remained bright in 2019 at the level of 4-8 keV surface brightness $\approx4\times10^{-13}$ erg s$^{-1}$ cm$^{-2}$ arcmin$^{-2}$ and the photon surface brightness in the 6.4 keV line of $\approx6\times10^{-6}$ photons s$^{-1}$ cm$^{-2}$ arcmin$^{-2}$. Significant morphological changes of the diffuse emission are clearly visible when compared to the earlier observations by \textit{Chandra} and \textit{XMM-Newton}. This justifies the Sgr~A complex as a primary target for the forthcoming high resolution imaging and polarimetric observations of this complex by \textit{Chandra} and \textit{IXPE}. 

{We detect hard diffuse emission featuring {the} 6.4 keV line from the two other regions adjacent to the Sgr~A complex, G0.17-0.08 and G0.17-0.17, from which comparable level of the line emission was observed before with \textit{XMM-Newton}. Further sensitive monitoring of this emission is crucial for the better characterisation and determination of its nature.}

We confirm low level of the {reflected} emission from once-bright Arches and Sgr~B2 clouds. Sensitive dedicated observations and data accumulated over long period of time might reveal dominance of multiply scattered emission coming from them at this stage. No brightening of other clouds to the level comparable to Sgr~A, Arches or Sgr~B2 in the past has been detected.

%------------------------------------------------

\section*{Data Availability}

The data of \textit{Chandra} and \textit{XMM-Newton} observatories are publicly available at the corresponding mission data archive. The \textit{SRG}/eROSITA data will become public in the course of the consolidated data release.

%------------------------------------------------
\section*{Acknowledgements}
\label{s:acknowledgements}
%------------------------------------------------

This work is based on observations with eROSITA telescope onboard \textit{SRG} space observatory. The \textit{SRG} observatory was built by Roskosmos in the interests of the Russian Academy of Sciences represented by its Space Research Institute (IKI) in the framework of the Russian Federal Space Program, with the participation of the Deutsches Zentrum für Luft- und Raumfahrt (DLR). The eROSITA X-ray telescope was built by a consortium of German Institutes led by MPE, and supported by DLR. The SRG spacecraft was designed, built, launched and is operated by the Lavochkin Association and its subcontractors. The science data are downlinked via the Deep Space Network Antennae in Bear Lakes, Ussurijsk, and Baikonur, funded by Roskosmos. The eROSITA data used in this work were converted to calibrated event lists using the eSASS software system developed by the German eROSITA Consortium and analysed using proprietary data reduction software developed by the Russian eROSITA Consortium.

The N$_2$H+ data used in this paper were obtained using the Mopra radio telescope, a part of the Australia Telescope National Facility which is funded by the Commonwealth of Australia for operation as a National Facility managed by CSIRO. The University of New South Wales (UNSW) digital filter bank (the UNSW-MOPS) used for the observations with Mopra was provided with support from the Australian Research Council (ARC), UNSW, Sydney and Monash Universities, as well as the CSIRO.

We thank Ekaterina Kuznetsova for providing us with the best fit model for the \textit{NuSTAR} data on the Arches Cluster. {We are grateful to the anonymous referee for helpful suggestions and comments.}

IK, EC and RS acknowledge partial support by the RSF grant 19-12-00369.

%------------------------------------------------
\bibliographystyle{mnras}
\bibliography{srge-gc-clouds.bib}
%------------------------------------------------

%\clearpage

\appendix
\label{a:mult}
%------------------------------------------------
\section{Reflection signal dominated by multiple scatterings}

{Let us consider a cloud with the moderate Thomson optical depth $\tau_{T}\sim0.1-0.5$, so that on the one hand the chance for a photon to be scattered two or more times is not infinitely small, while on the other hand the photon can cross the entire cloud without being absorbed.} When the illumination front from the primary source has already passed through the cloud, the intensity of the scattered radiation does not go to zero immediately. Indeed, there are photons which will be scattered more than once and, therefore, their arrival to an observer will be further delayed. For a compact cloud with a light-crossing time much shorter than the duration of the flare, this effect is not important and the observed spectrum will contain contributions of single and multiple scatterings in a correct proportion. However, for a larger cloud, whose light-crossing time is longer than the duration of the flare, this effect is significant. The lifetime of this delayed radiation is set by the light crossing time of the cloud itself or by the light crossing time of the scattering environment on larger scales,  e.g., low-density extended envelopes of molecular clouds. 

\begin{figure}
\includegraphics[bb=30 150 580 670, width=0.95\columnwidth]{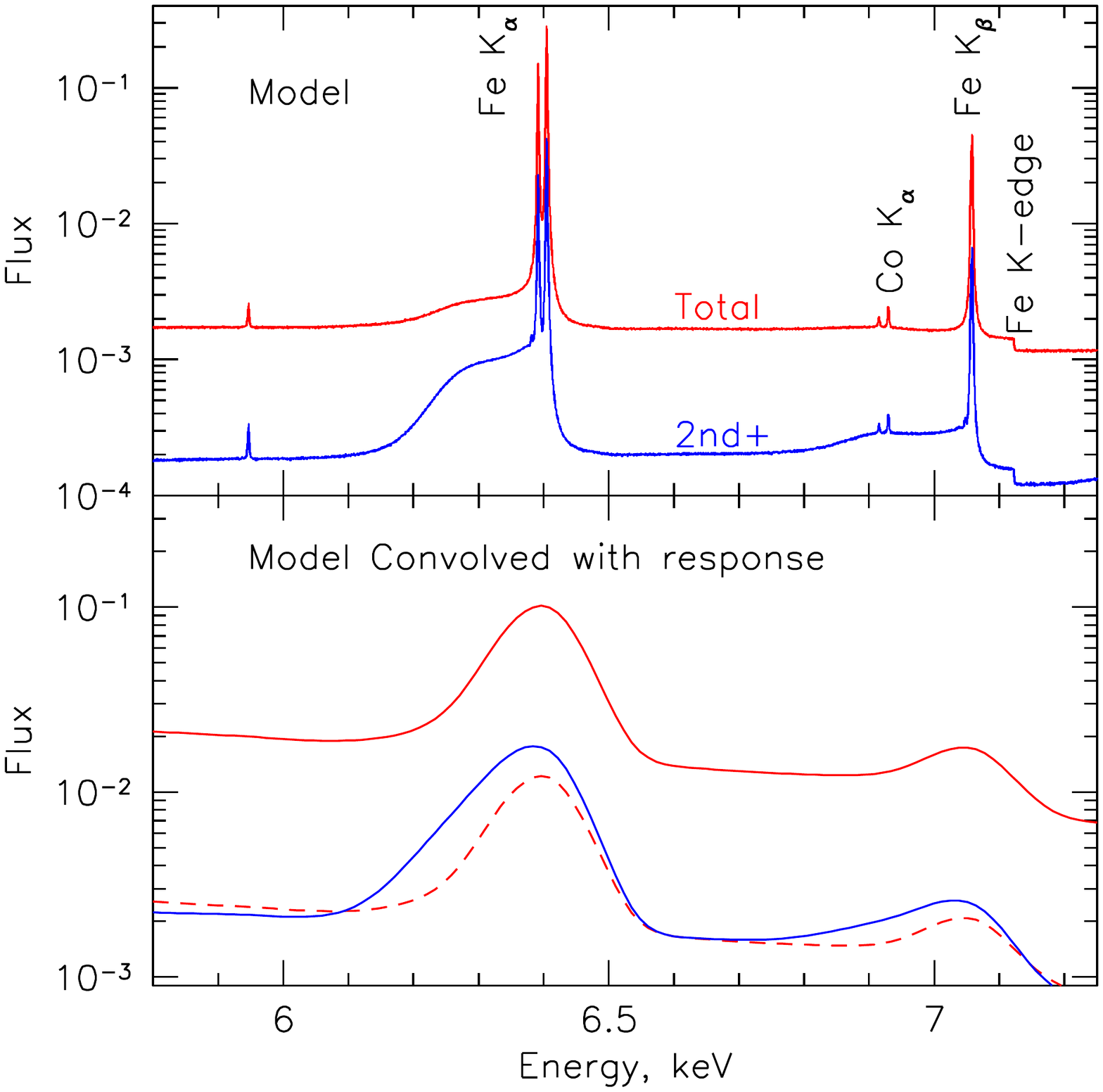}
\caption{ { {\bf Top panel:} Predicted spectrum (red line) arising from a slab of cold gas with the radial Thomson optical depth $\tau_T=0.2$ illuminated by a power-law X-ray continuum. Only a narrow region near the prominent iron 6.4 keV line is shown. 
For comparison, the blue line shows only photons that have experienced two interactions with the gas in the slab (either two Compton scatterings or a combination of the Compton scattering and fluorescence). This emission can be observed even after the primary photons have already left the cloud. In this Figure, the back-scattering peak of the Compton shoulder appears smooth due to scattering of photons by electrons bound hydrogen atoms. Free cold electrons would produce a sharp peak, while electrons hotter than $\sim 10$ eV would instead give rise to even smoother shape \citep{1996AstL...22..648S}. {\bf Bottom panel:} Solid red and blue lines show the spectra from the top panel convolved with the eROSITA spectral response. The dashed-red line shows the total spectrum scaled down to illustrate the changes in the relative intensity (i.e. equivalent width) and the shape of the line due to the increased contribution of the Compton shoulder.}
}
\label{fig:shoulder}
\end{figure}

 The delayed multiply-scattered emission will differ from the singly-scattered emission in intensity and spectral shape.  In particular, the equivalent width (EW) and shape of the fluorescent line (including the so called Compton shoulder) change \citep[see][for the discussion of the shoulder shape dependence on the temperature, ionization state and chemical composition of the medium]{1996AstL...22..648S,1998MNRAS.297.1279S,1998AstL...24..271V,2020MNRAS.495.1414K}. {For instance, when an X-ray photon is scattered by electrons bound in hydrogen or helium atoms, a fraction of photons will be scattered elastically (with no energy change), while another fraction will cause excitation to higher energy levels or ionization of the atom. In the latter case, the velocity distribution of electrons in the atom will contribute to the broadening of the emitted photon energy. Here, instead, we are considering an idealized problem, when the photon is scattered by free cold electrons. In this case, the change of the photon's energy is solely due to the recoil effect.} The example of the shape of the shoulder after the second scattering is shown in Fig.~\ref{fig:shoulder}.  This effect gives rise to another signature of the multiple scattering that can be found in the low resolution spectra -  fluorescent line centroid shift towards lower energies, since the EW grows mainly due to the Compton shoulder. 

\begin{figure}
\includegraphics[bb=30 150 580 670, width=0.95\columnwidth]{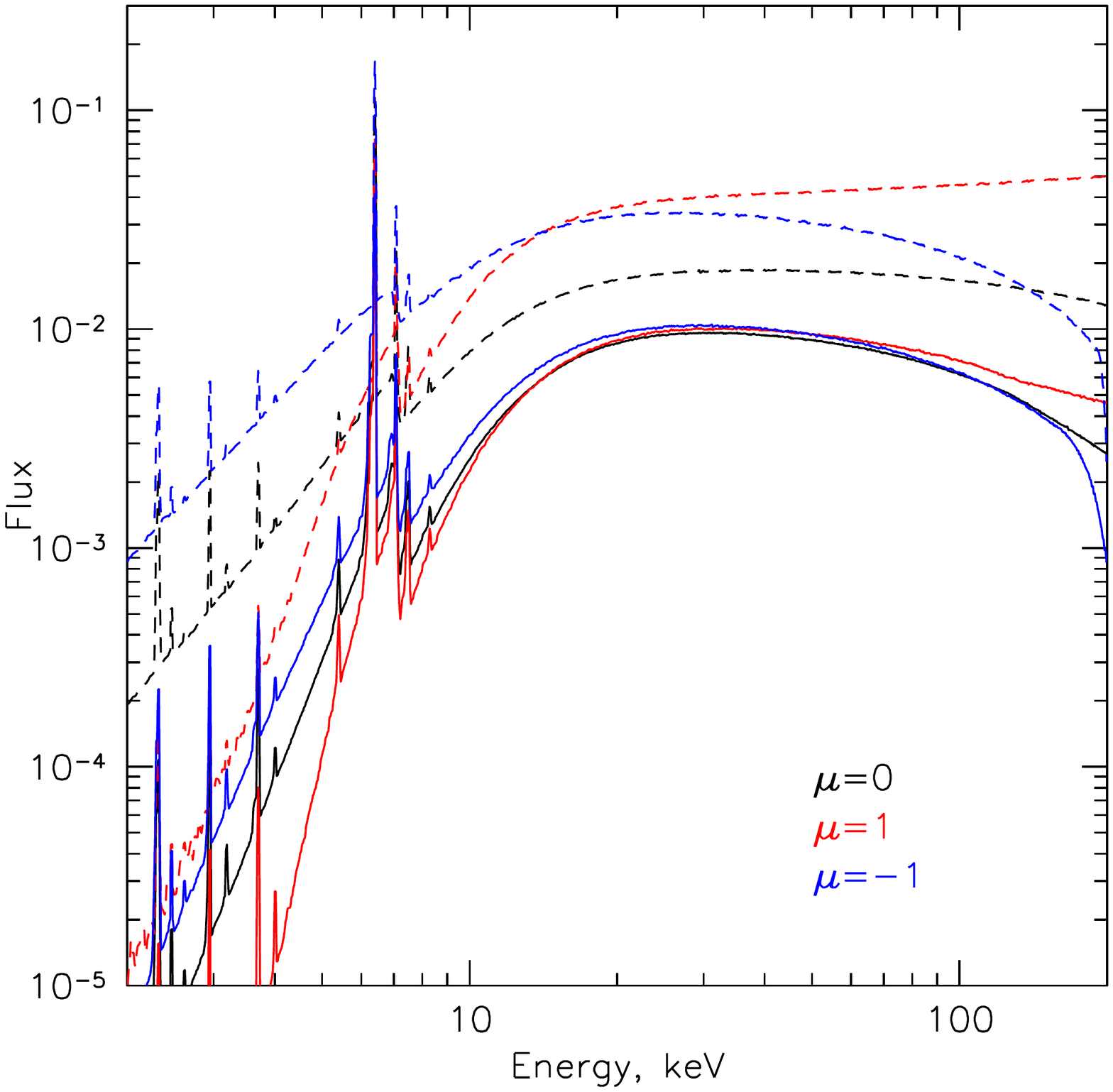}
\caption{Spectra of scattered emission emerging from a homogeneous spherical cloud with the Thomson optical depth $\tau_T=0.5$. Different colours correspond to different angles $\theta$ between the directions towards the primary source and the observer (as seen from the cloud). Here $\mu=\cos \theta=1$ corresponds to the cloud illuminated from the back (the emergent radiation after one scattering is dominated by forward scattering),  $\mu=0$ to the clouds illuminated from a side, and $\mu=-1$ to the cloud illuminated from the observer side. Dashed lines show the total scattered emission, including single- and mutiple-scattered photons. As expected, the scattering angles affect strongly both the low-energy side (due to photoelectric absorption) and the high-energy side (due to the angle-dependent recoil effect). The solid lines show the spectrum of photons experienced two or more "interactions" within the cloud. Unlike the dashed curves, the emergent spectra after more than one interaction show weaker dependence on the geometry.% of the problem.
\label{fig:scat_mul}
}
\end{figure}

Another interesting regime is relevant for energies higher than $\sim$20~keV, where the role of the recoil effects is very significant. The magnitude of the recoil effect directly depends on the scattering angle, being the largest for scattering by 180 degrees (back scattering) and zero for forward scattering. Because of this, the singly scattered spectrum should exhibit different level of steepening at high energies. Of course, this effect could be used to constrain the scattering angle too, provided that the shape of incident spectrum is known. After two (or more) scatterings, the distribution over scattering angles becomes much more uniform (once again, unless the cloud has a very asymmetric geometry). This implies that the "late" spectra should all have more or less similar shapes as shown in Fig. \ref{fig:scat_mul}. This means that the shape of the primary flare spectrum can be more robustly constraint above 20~keV observing scattered emission from clouds that have recently faded away, although the faintness of this emission complicates separation of the reflected component from various backgrounds and foregrounds.

The most promising for both of these kinds of studies is currently the Sgr~B2 cloud, which used to be a very bright source in the reflected emission before 2002, but has strongly faded away since then. Its large spatial extent implies that multiply-scattered photons should be present for decades, while its large mass should make the flux not negligibly small. The finite optical depth effects as well as non-uniform internal structure might be important for this cloud, but both of them might be readily accounted for in a dedicate Monte Carlo simulation \citep[e.g.][]{2016A&A...589A..88M}.

%%%%%%%%%%%%%%%%%%%%%%%%%%%%%%%%%%%%%%%%%%%%%%%%%%%%%%%%%%%%%%%%%

%%%%%%%%%%%%%%%%%%%%%%%%%%%%%%%%%%%%%%%%%%%%%%%%%%%%%%%%%%%%%%%%%

%\end{document}

%%%%%%%%%%%%%%%%%%%%%%%%%%%%%%%%%%%%%%%%%%%%%%%%%%%%%%%%%%%%%%%%%

% Don't change these lines
\bsp	% typesetting comment
\label{lastpage}
\end{document}